\documentclass[1pt]{article}
\usepackage[top=1.0in, bottom=1.0in, left=1.0in, right=1.0in]{geometry}
\usepackage{tikz}
\usepackage{braket}
\usepackage{amsmath}
\usetikzlibrary{arrows, positioning, automata, calc}
\usepackage{graphicx}
\usepackage{parskip}
\usepackage{titling}
\usepackage{qcircuit}
\usepackage{sectsty}
\usepackage{color}

\begin{document}
	\tikzset{vertex/.style = {shape=circle,draw,minimum size=2.0em}}
	\tikzset{root vertex/.style = {vertex, fill=black!50}}
	\tikzset{edge/.style = {->,> = latex',thick}}
	\tikzset{cross/.style = {pos=0.1}}
	\tikzset{line/.style={draw,thick,-latex', shorten >=2pt}}
	
	\tikzset{me/.style={to path={
\pgfextra{% 
 \pgfmathsetmacro{\startf}{-(#1-1)/2}  
 \pgfmathsetmacro{\endf}{-\startf} 
 \pgfmathsetmacro{\stepf}{\startf+1}}
 \ifnum 1=#1 -- (\tikztotarget)  \else
     let \p{mid}=($(\tikztostart)!0.5!(\tikztotarget)$) 
         in
\foreach \i in {\startf,\stepf,...,\endf}
    {%
     (\tikztostart) .. controls ($ (\p{mid})!\i*6pt!90:(\tikztotarget) $) .. (\tikztotarget)
      }
      \fi   
     \tikztonodes
}}}

\title{QuDot Nets: Quantum Computers and Bayesian Networks}
\author{Perry Sakkaris}
\date{\vspace{-5ex}}
\maketitle
\begin{abstract}
We present a new implementation of quantum computation that treats quantum computers as a special type of Bayesian Network called a QuDot Net. QuDot Nets allow for the efficient representation of some qubit systems. Single qubit quantum gates can be implemented as edge transformations on QuDot Nets. The $X$, $H$, $R(k)$, $M$ and SWAP gates are discussed in detail and results show linear scaling as the number of qubits are increased. We show that measurement and semi-quantum control gates can be efficiently implemented using QuDot Nets and present results from a QuDot Net implementation of the terminal Quantum Fourier Transform. We show how QuDot Nets can implement coherent control gates using multi-digraphs by labelling parallel edges. Lastly, we discuss implications to quantum foundations if a classical implementation of quantum computation is realized.
\end{abstract}

\section{Introduction}

Quantum computers can solve some computational problems exponentially faster than classical computers which has applications in the fields of cryptography [1], machine learning [2] and physics simulations [3]. These high profile applications of quantum computation have lead to a race to build new quantum computing hardware, but does quantum computation require new hardware? Quantum computation is a model of computation, much like Turing Machines are a model of computation. It is widely accepted but not proven that new hardware is needed to efficiently implement quantum computation. Similarly to how other open problems thought to be intractable seem to have an efficient classical implementation, such as Graph Isomorphism, an efficient implementation of  quantum computation on classical computers is a real possibility and we present a new method that takes us closer to this possibility.

We present a new graph theoretic implementation of the quantum computation model based on Bayesian Networks: QuDot Nets. QuDot Nets implement $n$-qubit states as nodes (or dots) on a directed, acyclic, weighted multi-graph (multi-digraph). The nodes are labeled $0$ or $1$ representing the possible values of a single qubit. An edge between nodes represents a tensor product and the edge weight represents a probability amplitude to observe the child node given the parent node. QuDot Nets are able to represent some $n$-qubit states using a linear amount of resources as opposed to the $2^n$ numbers required by a vector implementation. Single qubit quantum gates are implemented as transformations of edge weights thus allowing QuDot Nets to implement single qubit gates, measurements and semi-quantum control gates in linear time as opposed to requiring $2^n \times 2^n$ matrix multiplications. Coherent (non semi) two qubit control gates can be implemented by allowing multiple labeled edges to connect nodes making the QuDot Net a multi-digraph. The efficiency is still unresolved for these coherent control gates. The number of edges can still grow exponentially for coherent control gates, however, it remains to be seen if duplicate edges can be efficiently merged. 

QuDot Nets can also provide a deeper theoretical understanding of both quantum and classical computation. QuDot Nets allow us to analyze entanglement as a graph coloring problem. Also, if we are ultimately correct and quantum computation can be efficiently implemented on classical computers then $BPP = BQP$ and the strong Church-Turing Thesis is further confirmed. Furthermore, since there is strong evidence that $P = BPP$, if $BPP = BQP$ then it is possible that quantum mechanics can be de-randomized by a deterministic theory thus confirming Einstein's conviction that nature does not play dice.

\section{Representation and Measurement}

QuDot Nets are able to represent an $n$ qubit quantum state $\ket{q}$ if the quantum state starts in the ground state $\ket{000...0}$ and can be evolved to the desired state $\ket{q}$ using the quantum gates defined as edge transformations on QuDot Nets.  A QuDot Net represents a qubit as two graphical nodes side by side in a layer called a qubit layer. The number of qubit layers represents the total number of qubits, the topmost qubit layer is the first qubit and the $n$-th qubit layer is the $n$-th qubit. The first qubit layer is accessed by a special node called the root node.  The left node of a qubit layer is labeled the 0 node and the right node of the qubit layer is labeled the 1 node. The two nodes in each qubit layer are called siblings.  Directed, weighted edges go from a node in one qubit layer to a node in another qubit layer and represents a tensor product between those qubits. The weights of the directed edges are complex numbers and correspond to the probability amplitude to observe the child node given the parent node. There are $v = 2q + 1$ number of nodes to represent $q$ qubits and $e = 2v - 4$ edges. Therefore, the QuDot Net representation scales linearly in the number of qubits $q$ for qubit systems that are generated by starting in the ground state and evolved by using single qubit quantum gates defined on QuDot Nets ($H$, $R(k)$, $M$ and SWAP) and semi-quantum control gates. It remains to be seen if we can extend the linear (or polynomial) efficiency to coherent control gates.

\begin{figure}[!h]
\centering
\begin{tikzpicture}
	\node[root vertex] (r) at  (1.5,2) {};
	\node (rootLabel) at (1.5, 3) {$root$};
	
	\node[vertex] (00) at  (0,0) {$0$};
	\node[vertex] (01) at  (3,0) {$1$};	
	
	\node[vertex] (10) at  (0,-3) {$0$};
	\node[vertex] (11) at  (3,-3) {$1$};	
	
	\node[vertex] (20) at  (0,-6) {$0$};
	\node[vertex] (21) at  (3,-6) {$1$};	
	
	\node[vertex] (n0) at  (0,-9) {$0$};
	\node[vertex] (n1) at  (3,-9) {$1$};
	
	\draw[edge, cross] (r) edge node[left]{$\frac{1}{\sqrt{2}}$} (00);
	\draw[edge, cross] (r) edge node[right]{$\frac{1}{\sqrt{2}}$} (01);
	
	\draw[edge] (00) edge node[left]{$\frac{1}{\sqrt{2}}$}  (10);
	\draw[edge,cross] (00) edge node[right]{$\frac{1}{\sqrt{2}}$}  (11);
	
	\draw[edge, cross] (01) edge node[left]{$\frac{1}{\sqrt{2}}$} (10);
	\draw[edge] (01) edge node[right]{$\frac{1}{\sqrt{2}}$} (11);
	
	\draw[edge] (10) edge node[left]{$\frac{1}{\sqrt{2}}$} (20);
	\draw[edge, cross] (10) edge node[right]{$\frac{1}{\sqrt{2}}$} (21);
	
	\draw[edge, cross] (11) edge node[left]{$\frac{1}{\sqrt{2}}$} (20);
	\draw[edge] (11) edge node[right]{$\frac{1}{\sqrt{2}}$} (21);
	
	\draw[line, black, densely dashed] (-1,-1) rectangle (4,1);
	\node (q1) at (-2,0) {$qubit\phantom{a}1$};
	
	\draw[line, black, densely dashed] (-1,-4) rectangle (4,-2);
	\node (q2) at (-2,-3) {$qubit\phantom{a}2$};
	
	\draw[line, black, densely dashed] (-1,-7) rectangle (4,-5);
	\node (q3) at (-2,-6) {$qubit\phantom{a}3$};
	
	\draw[line, black, densely dashed] (-1,-10) rectangle (4,-8);
	\node (q4) at (-2,-9) {$qubit\phantom{a}n$};
	
	\path (20) to node {\vdots} (n0);
	\path (21) to node {\vdots} (n1);
\end{tikzpicture}
\caption{shows graphical QuDot Net representation of an $n$-qubit superposition}
\end{figure}
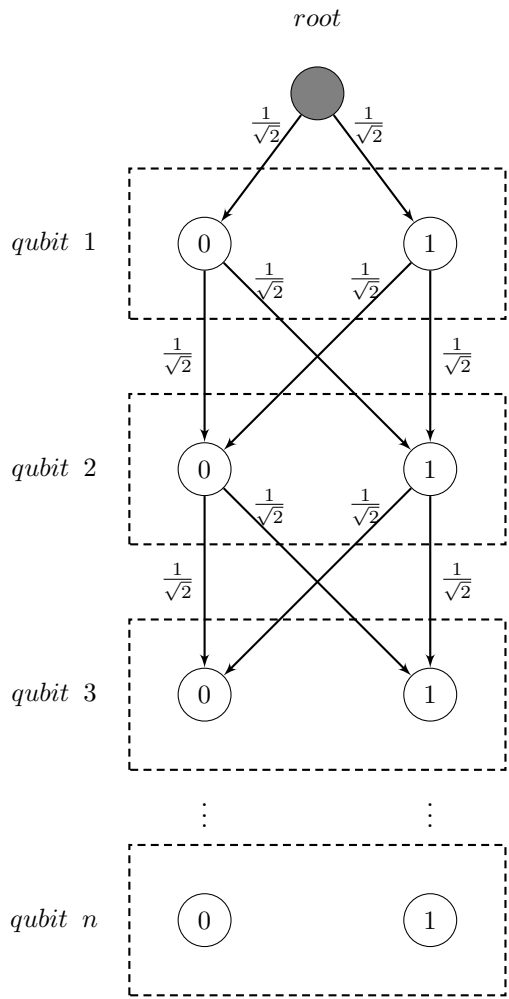

\newpage

Graph traversals of a QuDot Net correspond to quantum mechanical measurements in the computational basis. A possible traversal of the QuDot Net is a possible measurement outcome of the qubit state and the edge weights of the specific traversal multiplied together corresponds to the probability amplitude of that measurement. All possible traversals of the QuDot Net with a nonzero probability amplitude correspond to all possible measurement outcomes of the qubit state. A single traversal takes linear time and corresponds to a measurement in quantum mechanics. Obtaining all possible probability amplitudes requires the enumeration of all $2^n$ possible traversals of the QuDot Net. This corresponds to the fact that Nature ``hides'' the probability amplitudes, therefore, QuDot Nets are consistent with the framework of quantum mechanics.

Consider $\ket{q} = \frac{1}{\sqrt{8}} \left( \ket{000} + \ket{001} + \ket{010} + \ket{011} + \ket{100} + \ket{101} + \ket{110} + \ket{111}  \right)$

\begin{figure}[!h]
\centering
\begin{tikzpicture}
	\node[root vertex] (r) at  (1.5,2) {};
	
	\node[vertex] (00) at  (0,0) {$0$};
	\node[vertex] (01) at  (3,0) {$1$};	
	
	\node[vertex] (10) at  (0,-3) {$0$};
	\node[vertex] (11) at  (3,-3) {$1$};	
	
	\node[vertex] (20) at  (0,-6) {$0$};
	\node[vertex] (21) at  (3,-6) {$1$};

	\draw[edge, cross] (r) edge node[left]{$\frac{1}{\sqrt{2}}$} (00);
	\draw[edge, cross] (r) edge node[right]{$\frac{1}{\sqrt{2}}$} (01);
	
	\draw[edge] (00) edge node[left]{$\frac{1}{\sqrt{2}}$}  (10);
	\draw[edge,cross] (00) edge node[right]{$\frac{1}{\sqrt{2}}$}  (11);
	
	\draw[edge, cross] (01) edge node[left]{$\frac{1}{\sqrt{2}}$} (10);
	\draw[edge] (01) edge node[right]{$\frac{1}{\sqrt{2}}$} (11);
	
	\draw[edge] (10) edge node[left]{$\frac{1}{\sqrt{2}}$} (20);
	\draw[edge, cross] (10) edge node[right]{$\frac{1}{\sqrt{2}}$} (21);
	
	\draw[edge, cross] (11) edge node[left]{$\frac{1}{\sqrt{2}}$} (20);
	\draw[edge] (11) edge node[right]{$\frac{1}{\sqrt{2}}$} (21);
	
\end{tikzpicture}
\caption{shows QuDot Net of $\ket{q}$}
\end{figure}
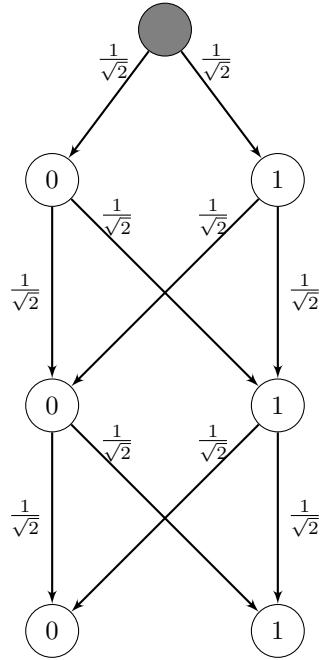

\paragraph{}
There are eight possible measurement outcomes of $\ket{q}$ corresponding to the possible traversals of the QuDot Net in figure 2:
\begin{enumerate}
	\item{ $\ket{000}$ with probability amplitude $\frac{1}{\sqrt{2}} \times \frac{1}{\sqrt{2}} \times \frac{1}{\sqrt{2}} = \frac{1}{2\sqrt{2}}$ }
	\item{ $\ket{001}$ with probability amplitude $\frac{1}{\sqrt{2}} \times \frac{1}{\sqrt{2}} \times \frac{1}{\sqrt{2}} = \frac{1}{2\sqrt{2}}$}
	\item{ $\ket{010}$ with probability amplitude $\frac{1}{\sqrt{2}} \times \frac{1}{\sqrt{2}} \times \frac{1}{\sqrt{2}} = \frac{1}{2\sqrt{2}}$ }
	\item{ $\ket{011}$ with probability amplitude $\frac{1}{\sqrt{2}} \times \frac{1}{\sqrt{2}} \times \frac{1}{\sqrt{2}} = \frac{1}{2\sqrt{2}}$ }
	
	\item{ $\ket{100}$ with probability amplitude $\frac{1}{\sqrt{2}} \times \frac{1}{\sqrt{2}} \times \frac{1}{\sqrt{2}} = \frac{1}{2\sqrt{2}}$ }
	\item{ $\ket{101}$ with probability amplitude $\frac{1}{\sqrt{2}} \times \frac{1}{\sqrt{2}} \times \frac{1}{\sqrt{2}} = \frac{1}{2\sqrt{2}}$ }
	\item{ $\ket{110}$ with probability amplitude $\frac{1}{\sqrt{2}} \times \frac{1}{\sqrt{2}} \times \frac{1}{\sqrt{2}} = \frac{1}{2\sqrt{2}}$ }
	\item{ $\ket{111}$ with probability amplitude $\frac{1}{\sqrt{2}} \times \frac{1}{\sqrt{2}} \times \frac{1}{\sqrt{2}} = \frac{1}{2\sqrt{2}}$ }
\end{enumerate}

\paragraph{}
Consider $\ket{g} = \ket{000}$

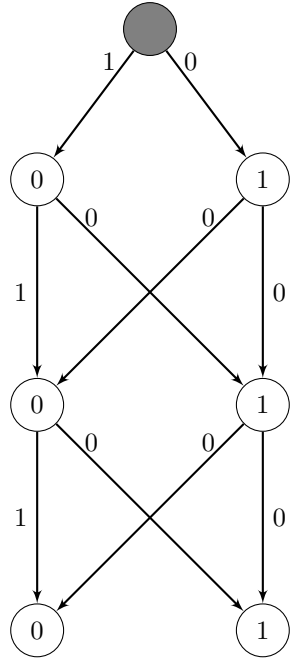
\begin{figure}[!h]
\centering
\begin{tikzpicture}
	\node[root vertex] (r) at  (1.5,2) {};
	
	\node[vertex] (00) at  (0,0) {$0$};
	\node[vertex] (01) at  (3,0) {$1$};	
	
	\node[vertex] (10) at  (0,-3) {$0$};
	\node[vertex] (11) at  (3,-3) {$1$};	
	
	\node[vertex] (20) at  (0,-6) {$0$};
	\node[vertex] (21) at  (3,-6) {$1$};

	\draw[edge, cross] (r) edge node[left]{$1$} (00);
	\draw[edge, cross] (r) edge node[right]{$0$} (01);
	
	\draw[edge] (00) edge node[left]{$1$}  (10);
	\draw[edge,cross] (00) edge node[right]{$0$}  (11);
	
	\draw[edge, cross] (01) edge node[left]{$0$} (10);
	\draw[edge] (01) edge node[right]{$0$} (11);
	
	\draw[edge] (10) edge node[left]{$1$} (20);
	\draw[edge, cross] (10) edge node[right]{$0$} (21);
	
	\draw[edge, cross] (11) edge node[left]{$0$} (20);
	\draw[edge] (11) edge node[right]{$0$} (21);
	
\end{tikzpicture}
\caption{shows QuDot Net of $\ket{g}$}
\end{figure}

\paragraph{}
There is only one possible measurement outcome of $\ket{g}$ corresponding to the one nonzero traversal of the QuDot Net in figure 3:

\begin{enumerate}
	\item{$\ket{000}$ with probability amplitude $1 \times 1 \times 1 = 1$}
\end{enumerate}

\textit{QuDot Net Traversal Algorithm}: begin on root, while a node has child edges let $p_0$, $p_1$ be the edge weights of the child edges directed towards the $0$ and $1$ child nodes respectively. Generate a random number $r \in [0,1]$. If $r \le \left| p_0 \right| ^2$ then visit the $0$ child node, else visit the $1$ child node. The labels of the nodes visited is the measured state. This algorithm will reproduce the probability distribution of the qubit state and is in $O(n)$ where $n$ is the number of qubits. QuDot Nets implement measurements in linear time.

\section{Singe Qubit Gates}

\subsection{Quantum gates as edge weight transformations}
QuDot Nets allow for the efficient implementation of some single qubit gates by treating them as operations on the edge weights. The unitary transformations that the quantum circuit model uses as quantum gates are replaced with transformations of edge weights in a qubit layer. QuDot Net implementations of single qubit gates define specific rules on how to transform parent and child edge weights of sibling nodes in a qubit layer as outlined in figure 4. The $X$, $H$, $M$ and $R(k)$ single qubit gates are described in detail, however, anyone of ordinary skill in the methods of QuDot Nets will appreciate that many variations and alterations to the following details may be used to create additional single qubit gates. We implement our methods for the gates we define using the Java programming language and show results for the scalability of QuDot Net defined quantum gates. A clear linear scale is shown for all defined gates.

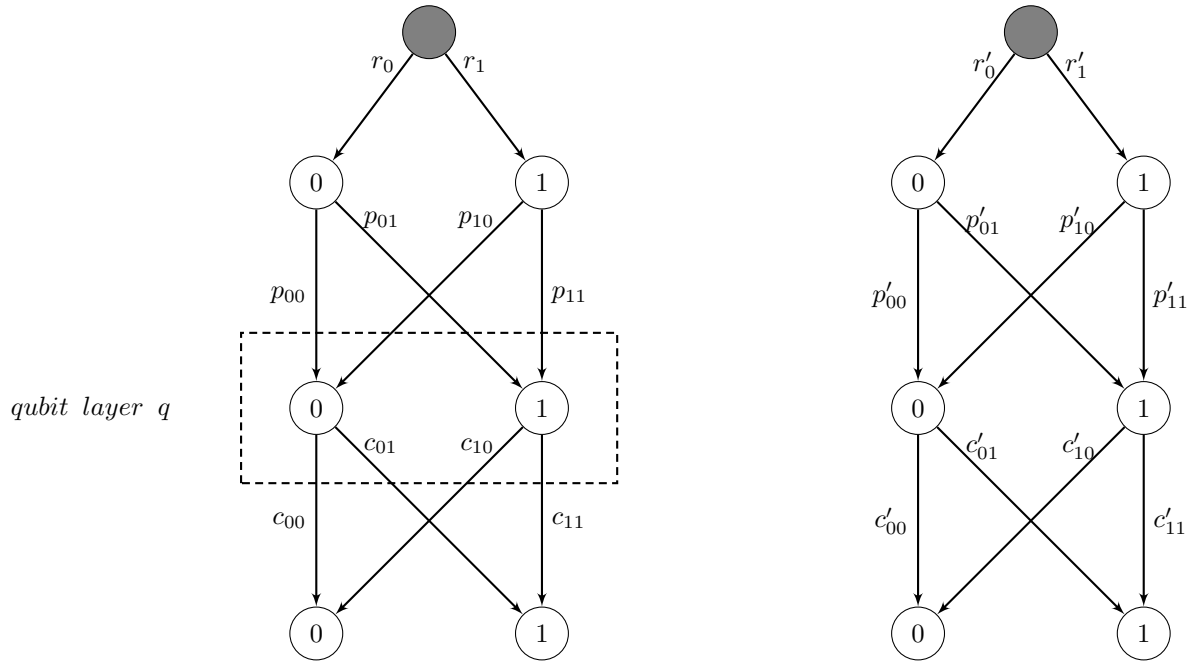
\begin{figure}[h]
\centering
\begin{tikzpicture}
	
	\node[root vertex] (r) at  (1.5,2) {};
	\node[vertex] (00) at  (0,0) {$0$};
	\node[vertex] (01) at  (3,0) {$1$};	
	
	\node[vertex] (10) at  (0,-3) {$0$};
	\node[vertex] (11) at  (3,-3) {$1$};	
	
	\node[vertex] (20) at  (0,-6) {$0$};
	\node[vertex] (21) at  (3,-6) {$1$};

	\draw[edge] (00) edge node[left]{$p_{00}$}  (10);
	\draw[edge,cross] (00) edge node[right]{$p_{01}$}  (11);
	
	\draw[edge, cross] (01) edge node[left]{$p_{10}$} (10);
	\draw[edge] (01) edge node[right]{$p_{11}$} (11);
	
	\draw[edge] (10) edge node[left]{$c_{00}$} (20);
	\draw[edge, cross] (10) edge node[right]{$c_{01}$} (21);
	
	\draw[edge, cross] (11) edge node[left]{$c_{10}$} (20);
	\draw[edge] (11) edge node[right]{$c_{11}$} (21);
	
	\draw[edge, cross] (r) edge node[left]{$r_0$} (00);
	\draw[edge, cross] (r) edge node[right]{$r_1$} (01);
	
	\draw[line, black, densely dashed] (-1,-4) rectangle (4,-2);
	\node (q2) at (-3,-3) {$qubit\phantom{a}layer\phantom{a}q$};
	
	\node[root vertex] (r2) at  (9.5,2) {};
	\node[vertex] (00) at  (8,0) {$0$};
	\node[vertex] (01) at  (11,0) {$1$};	
	
	\node[vertex] (10) at  (8,-3) {$0$};
	\node[vertex] (11) at  (11,-3) {$1$};	
	
	\node[vertex] (20) at  (8,-6) {$0$};
	\node[vertex] (21) at  (11,-6) {$1$};

	\draw[edge] (00) edge node[left]{$p'_{00}$}  (10);
	\draw[edge,cross] (00) edge node[right]{$p'_{01}$}  (11);
	
	\draw[edge, cross] (01) edge node[left]{$p'_{10}$} (10);
	\draw[edge] (01) edge node[right]{$p'_{11}$} (11);
	
	\draw[edge] (10) edge node[left]{$c'_{00}$} (20);
	\draw[edge, cross] (10) edge node[right]{$c'_{01}$} (21);
	
	\draw[edge, cross] (r2) edge node[left]{$r'_0$} (00);
	\draw[edge, cross] (r2) edge node[right]{$r'_1$} (01);
	
	\draw[edge, cross] (11) edge node[left]{$c'_{10}$} (20);
	\draw[edge] (11) edge node[right]{$c'_{11}$} (21);	
	
\end{tikzpicture}
\caption{single qubit gate applied on qubit $q$}
\end{figure}

Note that special care must be taken for the first qubit because the parent node for the first qubit is the \textit{root} node which is not another qubit node. The edges of the root node are labeled $r_0$ and $r_1$ for the edges directed towards the zero node and one node respectively. We show a separate description of the first qubit in our gates below for added clarity.

\subsection{X: NOT gate}

To apply the $X$ gate to qubit $q$ in a QuDot Net:

    \begin{enumerate}
	\item go to qubit layer $q$
	
	\item swap child edge weights of siblings in qubit layer $q$:
	\begin{align*}
	    c'_{00} &= c_{10}  \\
             c'_{01} &= c_{11} \\
	   c'_{10} &= c_{00} \\
	   c'_{11} &= c_{01}
	\end{align*}

	\item if $q=1$ (root node parent):
		\begin{align*}
			r'_0 = r_1 \\
			r'_1 = r_0
		\end{align*}	
	\item if $q \ne 1$ swap parent edge weights of siblings in qubit layer $q$: 
		\begin{align*}
			p'_{00} &= p_{01} \\
			p'_{10} &= p_{11} \\
			p'_{01} &= p_{00} \\
			p'_{11} &= p_{10}
		\end{align*}
    \end{enumerate}

\begin{figure}[!ht]
\includegraphics[scale=0.5]{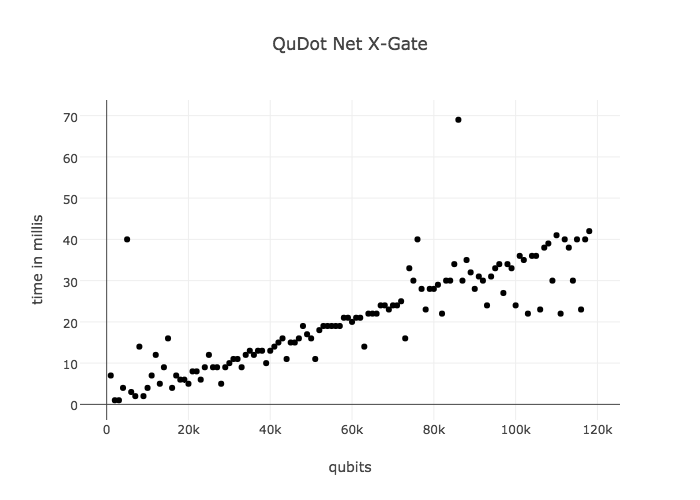}
\centering
\caption{shows linear scaling of $X$ gate in a QuDot Net up to 118,000 qubits}
\end{figure}

\newpage
\subsection{H: Haddamard gate}

To apply the $H$ gate to qubit $q$ in a QuDot Net:
    \begin{enumerate}
	\item go to qubit layer $q$
	
		\item if $q=1$ (root node parent):
		\begin{align*}
			r'_0 &= r_{0} \left( \frac{1}{\sqrt{2}} \right) + r_{1}  \left( \frac{1}{\sqrt{2}} \right) \\ 
			r'_1 &= r_{0} \left( \frac{1}{\sqrt{2}} \right) + r_{1}  \left( \frac{-1}{\sqrt{2}} \right) 
		\end{align*}	
		
	\item if $q \ne 1$ interfere parent edges in qubit layer $q$:
		\begin{align*}
			p'_{00} &= p_{00} \left( \frac{1}{\sqrt{2}} \right) + p_{01}  \left( \frac{1}{\sqrt{2}} \right) \\
			p'_{10} &=  p_{10} \left( \frac{1}{\sqrt{2}} \right) + p_{11}  \left( \frac{1}{\sqrt{2}} \right)\\
			p'_{01} &=  p_{00} \left( \frac{1}{\sqrt{2}} \right) + p_{01}  \left( \frac{-1}{\sqrt{2}} \right) \\
			p'_{11} &=  p_{10} \left( \frac{1}{\sqrt{2}} \right) + p_{11}  \left( \frac{-1}{\sqrt{2}} \right)		
		\end{align*}
		
	\item copy nonzero child edges: 
		\begin{itemize}
			\item if $c_{00} = 0$ and $c_{10} \ne 0$ or $c_{00}  \ne 0$ and $c_{10} = 0$ then set $c_{00} = c_{10}$
			\item if $c_{01} = 0$ and $c_{11} \ne 0$ or $c_{01}  \ne 0$ and $c_{11} = 0$ then set $c_{01} = c_{11}$
		\end{itemize}
    \end{enumerate}

\begin{figure}[!h]
\includegraphics[scale=0.4]{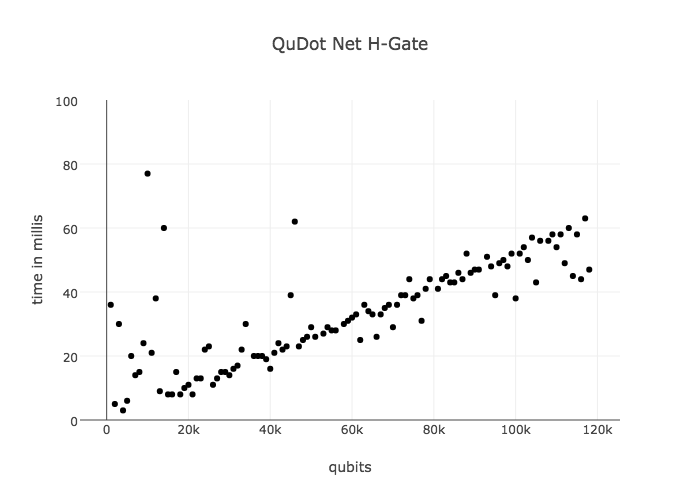}
\centering
\caption{shows linear scaling of $H$ gate in a QuDot Net up to 118,000 qubits}
\end{figure}

\newpage
\subsection{R(k): phase gates}

$R(k)$ gates are the family of phase gates. For each $k \ge 0$ we have a distinct gate $R(k)$. The following method applies to the family of phase gates:

$$
R(k) = 
\begin{bmatrix}
	1 & 0 \\
	0 & \phi(k)
\end{bmatrix}
$$

where $\phi(k)$ is the \textit{phase} defined as $$\phi(k) = e^{\frac{2 \pi i}{2^{k}}}$$

To apply the $R(k)$ gate to qubit number $q$ in a QuDot Net: 
    \begin{enumerate}
	\item go to qubit layer $q$
	
		\item if $q=1$ (root node parent):
		\begin{align*}
			r'_1 = r_1 * \phi(k)
		\end{align*}	
		
	\item if $q \ne 1$ apply the phase to the parent edges of the 1 node:
		\begin{align*}
			p'_{01} &= p_{01} * \phi(k) \\
			p'_{11} &= p_{11} * \phi(k)
		\end{align*}
    \end{enumerate}

\begin{figure}[!h]
\includegraphics[scale=0.5]{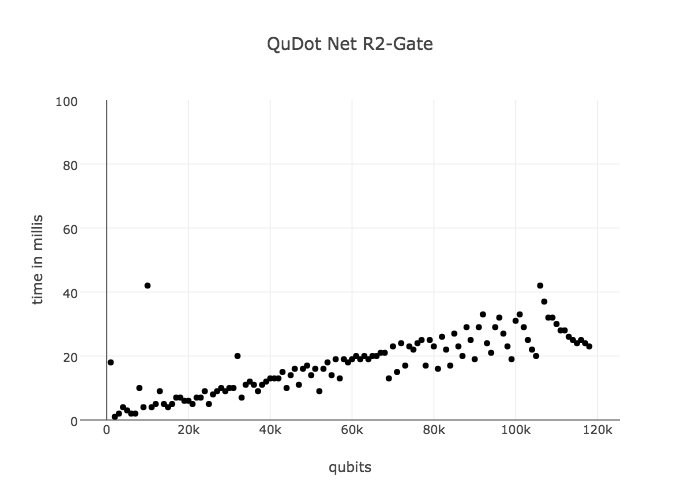}
\centering
\caption{shows linear scaling of $R(2)$ gate in a QuDot Net up to 118,000 qubits}
\end{figure}

\newpage

\subsection{Accuracy of Single Qubit Gates}

We tested the accuracy of the single qubit gates $X$, $H$ and $R(k)$ by directly comparing 10 qubit states to a matrix mechanics emulation. We use the open source Python library \textit{QuDotPy} for the matrix mechanics emulation.  Let $\rho$ be the density matrix of the QuDot Net quantum state and $\sigma$ be the density matrix of the exact state obtained via matrix mechanics emulation. We use the Fidelity Measure to compare the output:
$$ F(\rho, \sigma) = Tr\sqrt{\rho^{1/2}\sigma\rho^{1/2}} $$

Our testing process starts in the ground state and applies a gate to: entire state, all even qubits, subsets of even qubits, odd qubits, subset of odd qubits; finally, repeat with a state in a complete superposition. For each test we run the operation through both QuDot Nets and a classical emulation and compare the results to obtain the fidelity. We use the Worst Fidelity measurement as our fidelity for a gate. Our Fidelity results are:

\begin{enumerate}
	\item $X$ Gate: 0.99999999999999986
	\item $H$ Gate: 0.99999999999999730
	\item $R(k)$ Gates: 09999998801650631
\end{enumerate}

\subsection{M: measurment gate}
QuDot Nets allow for the efficient implementation of the single qubit measurement gate $M$. The single qubit measurement gate $M$ assigns a definite value to the qubit $q$ of $0$ or $1$. To apply the single qubit measurement gate $M$ to the qubit q:

\begin{enumerate}
	\item go to qubit layer $q$
	
	\item if the $0$ node is the only node with parent edges that have nonzero weights then $q = 0$
	
	\item if the $1$ node is the only node with parent edges that have nonzero weights then $q = 1$
	
	\item if both the $0$ and $1$ nodes have parent edges with nonzero weights then calculate the probability of the 0 node, defined as $P(0)$, then generate a random number between 0 and 1 defined as $r$. If  $r <= P(0)$ then $q = 0$, the $0$ node is activated and the $1$ node is deactivated. If $r > P(0)$ then $q = 1$, the $1$ node is activated and the $0$ node is deactivated.
	
	\begin{enumerate}
	\item To calculate $P(0)$: the 0 node has at most two nonzero parent edges. Let $w_1$, $w_2$ be the edge weights of those two parents. Let $c$ be the number of edge weights that are nonzero. Then $P(0) = \frac{\left| w_1 + w_2 \right|^2}{c^2}$
	
	\item To deactivate a node: set the weight of all parent edges to $0$
	
	\item To activate a node: if the node has $2$ nonzero parent edges then multiply the parent edges by $\sqrt{2}$ otherwise leave edge weights the same
	
	\end{enumerate}
\end{enumerate}

\section{Quantum Circuits}

QuDot Nets are able to efficiently implement quantum circuits composed of single qubit gates. We always begin in the ground state $\ket{000...0}$ and apply quantum gates one at a time to the QuDot Net as they appear in our quantum circuits. Then we traverse the QuDot Net using the traversal algorithm and that traversal is our state measurement. We illustrate the procedure with a simple example. Consider the quantum circuit in figure 8:

\begin{figure}[!h]
\centering
\mbox{
\Qcircuit @C=1em @R=.7em {
	\lstick{\ket{0}} &  \gate{H} & \qw & \qw & \qw \\
	\lstick{\ket{0}} & \qw & \qw & \gate{H} & \qw  \\
	\lstick{\ket{0}} & \qw & \gate{X}  & \qw & \qw
}
}
\caption{shows a simple quantum circuit used to explain QuDot Nets}
\end{figure}
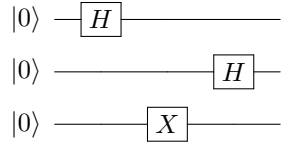 

We initialize our QuDot Net in the ground state $\ket{q}$ shown in figure 9. Note that edges having a weight of $0$ have been omitted from the drawing of the QuDot Net to make the graph easier to follow for the reader. 

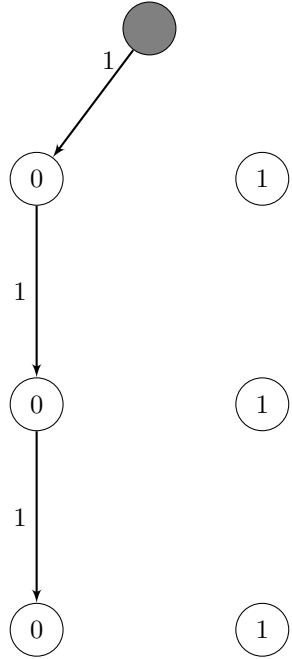
\begin{figure}[!h]
\centering
\begin{tikzpicture}
	\node[root vertex] (r) at  (1.5,2) {};
	
	\node[vertex] (00) at  (0,0) {$0$};
	\node[vertex] (01) at  (3,0) {$1$};	
	
	\node[vertex] (10) at  (0,-3) {$0$};
	\node[vertex] (11) at  (3,-3) {$1$};	
	
	\node[vertex] (20) at  (0,-6) {$0$};
	\node[vertex] (21) at  (3,-6) {$1$};

	\draw[edge, cross] (r) edge node[left]{$1$} (00);
	
	\draw[edge] (00) edge node[left]{$1$}  (10);	
	
	\draw[edge] (10) edge node[left]{$1$} (20);

\end{tikzpicture}
\caption{shows QuDot Net of $\ket{q}= \ket{000}$}
\end{figure}

\newpage
Apply $H$ to qubit 1:

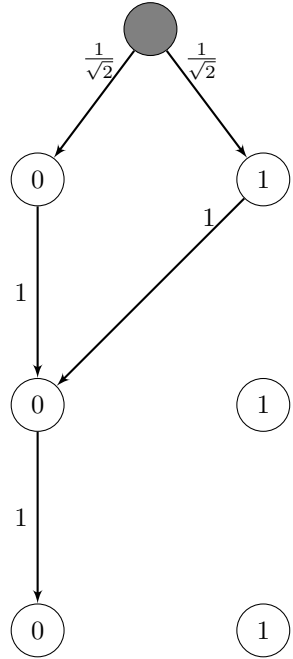
\begin{figure}[!h]
\centering
\begin{tikzpicture}
	\node[root vertex] (r) at  (1.5,2) {};
	
	\node[vertex] (00) at  (0,0) {$0$};
	\node[vertex] (01) at  (3,0) {$1$};	
	
	\node[vertex] (10) at  (0,-3) {$0$};
	\node[vertex] (11) at  (3,-3) {$1$};	
	
	\node[vertex] (20) at  (0,-6) {$0$};
	\node[vertex] (21) at  (3,-6) {$1$};

	\draw[edge, cross] (r) edge node[left]{$\frac{1}{\sqrt{2}}$} (00);
	\draw[edge, cross] (r) edge node[right]{$\frac{1}{\sqrt{2}}$} (01);
	
	\draw[edge] (00) edge node[left]{$1$}  (10);
	\draw[edge, cross] (01) edge node[left]{$1$} (10);	
	
	\draw[edge] (10) edge node[left]{$1$} (20);

\end{tikzpicture}
\caption{shows QuDot Net of $\ket{q}= \frac{1}{\sqrt{2}} \left( \ket{000} + \ket{100} \right)$}
\end{figure}

Apply $X$ to qubit 3:
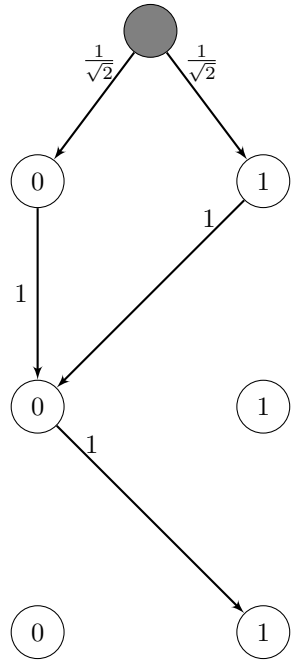
\begin{figure}[!ht]
\centering
\begin{tikzpicture}
	\node[root vertex] (r) at  (1.5,2) {};
	
	\node[vertex] (00) at  (0,0) {$0$};
	\node[vertex] (01) at  (3,0) {$1$};	
	
	\node[vertex] (10) at  (0,-3) {$0$};
	\node[vertex] (11) at  (3,-3) {$1$};	
	
	\node[vertex] (20) at  (0,-6) {$0$};
	\node[vertex] (21) at  (3,-6) {$1$};

	\draw[edge, cross] (r) edge node[left]{$\frac{1}{\sqrt{2}}$} (00);
	\draw[edge, cross] (r) edge node[right]{$\frac{1}{\sqrt{2}}$} (01);
	
	\draw[edge] (00) edge node[left]{$1$}  (10);
	\draw[edge, cross] (01) edge node[left]{$1$} (10);	
	
	\draw[edge, cross] (10) edge node[right]{$1$} (21);

\end{tikzpicture}
\caption{shows QuDot Net of $\ket{q}= \frac{1}{\sqrt{2}} \left( \ket{001} + \ket{101} \right)$}
\end{figure}

\newpage
Apply $H$ to qubit 2:
\begin{figure}[!ht]
\centering
\begin{tikzpicture}
	\node[root vertex] (r) at  (1.5,2) {};
	
	\node[vertex] (00) at  (0,0) {$0$};
	\node[vertex] (01) at  (3,0) {$1$};	
	
	\node[vertex] (10) at  (0,-3) {$0$};
	\node[vertex] (11) at  (3,-3) {$1$};	
	
	\node[vertex] (20) at  (0,-6) {$0$};
	\node[vertex] (21) at  (3,-6) {$1$};

	\draw[edge, cross] (r) edge node[left]{$\frac{1}{\sqrt{2}}$} (00);
	\draw[edge, cross] (r) edge node[right]{$\frac{1}{\sqrt{2}}$} (01);
	
	\draw[edge] (00) edge node[left]{$\frac{1}{\sqrt{2}}$}  (10);
	\draw[edge,cross] (00) edge node[right]{$\frac{1}{\sqrt{2}}$}  (11);
	
	\draw[edge, cross] (01) edge node[left]{$\frac{1}{\sqrt{2}}$} (10);	
	\draw[edge, cross] (01) edge node[right] {$\frac{1}{\sqrt{2}}$} (11);
	
	\draw[edge, cross] (10) edge node[right]{$1$} (21);
	\draw[edge, cross] (11) edge node[right] {$1$} (21);

\end{tikzpicture}
\caption{shows QuDot Net of $\ket{q}= \frac{1}{2} \left( \ket{001} + \ket{011} +  \ket{101} + \ket{111}  \right)$}
\end{figure}
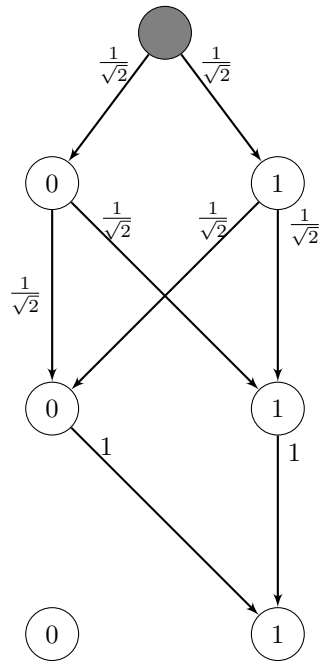

If we now run our QuDot Net traversal algorithm we will perform a measurement thus concluding the implementation of the quantum circuit in figure 8. If we perform the QuDot Net traversal algorithm many times we will obtain different traversal corresponding to the different possible measurement of the state $\ket{q}$. The probability distribution we obtain from performing many traversals of the QuDot Net in figure 12 matches the probability distribution of $\ket{q}$, for example, we will obtain the traversal \textit{011} with a probability of $\frac{1}{4}$. 

\section{Semi-Quantum Computation}

\subsection{semi-quantum control gates}
QuDot Nets can be used to efficiently implement semi-quantum circuits on classical hardware. A semi-quantum circuit is defined by how a control-$U$ operation (where $U$ is a quantum gate) is handled. If your method measures the control qubit and uses the result of the measurement to decide whether the $U$ gate should be applied to the target qubit; then it is called a semi-quantum gate and is an application of the \textit{Principle of Deffered Measurement} depicted in figure 13.

\begin{figure}[!ht]
\centering
\mbox{
\Qcircuit @C=1em @R=.7em {
	& \ctrl{1} & \meter & \cw  & \push{\rule{.3em}{0em}=\rule{.3em}{0em}}  & \meter & \control \cw & \push{\rule{.3em}{0em}=\rule{.3em}{0em}}              & & \control & \cw  \\
         & \gate{U} & \qw & \qw    & \push{\rule{.3em}{0em}\phantom{=}\rule{.3em}{0em}} & \qw &  \gate{U} \cwx      & & &  \gate{U} \cwx & \qw
}
}
\caption{Principle of Deferred Measurement and semi-quantum control-$U$ gate}
\end{figure}
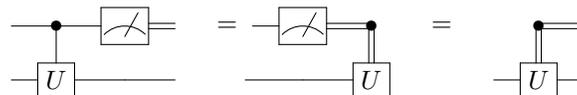

After a measurement is performed on the control qubit then it collapses to a definite value and further interference cannot occur. The control qubit becomes a classical bit after measurement, therefore, the only operations allowed after a semi-quantum control gate on the control qubit are other measurements or controls.  Circuits that are composed entirely of single qubit gates, semi-quantum control gates and treat the control qubit classically after a control are called \textit{terminal} or \textit{semi-quantum} circuits. One such circuit is the Quantum Fourier Transform (QFT) which we implement using QuDot Nets. Semi-quantum control gates are implemented with QuDot Nets by measuring the control qubit using the $M$ gate and applying the $U$ single qubit gate on the target qubit if the measurement result is $1$.

\subsection{SWAP gate}

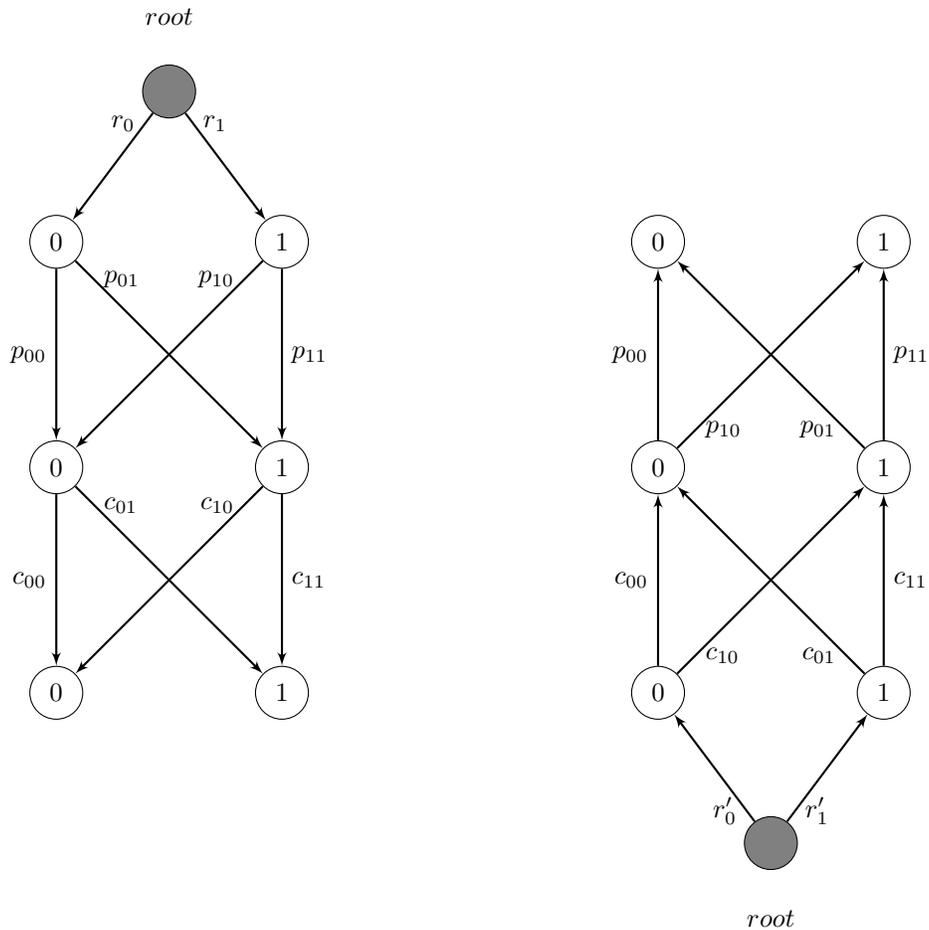
\begin{figure}[!h]
\centering
\begin{tikzpicture}
	
	\node[root vertex] (r) at  (1.5,2) {};
	\node (rootLabel) at (1.5, 3) {$root$};
	
	\node[vertex] (00) at  (0,0) {$0$};
	\node[vertex] (01) at  (3,0) {$1$};	
	
	\node[vertex] (10) at  (0,-3) {$0$};
	\node[vertex] (11) at  (3,-3) {$1$};	
	
	\node[vertex] (20) at  (0,-6) {$0$};
	\node[vertex] (21) at  (3,-6) {$1$};

	\draw[edge, cross] (r) edge node[left]{$r_0$} (00);
	\draw[edge, cross] (r) edge node[right]{$r_1$} (01);
	
	\draw[edge] (00) edge node[left]{$p_{00}$}  (10);
	\draw[edge,cross] (00) edge node[right]{$p_{01}$}  (11);
	
	\draw[edge, cross] (01) edge node[left]{$p_{10}$} (10);
	\draw[edge] (01) edge node[right]{$p_{11}$} (11);
	
	\draw[edge] (10) edge node[left]{$c_{00}$} (20);
	\draw[edge, cross] (10) edge node[right]{$c_{01}$} (21);
	
	\draw[edge, cross] (11) edge node[left]{$c_{10}$} (20);
	\draw[edge] (11) edge node[right]{$c_{11}$} (21);

	\node[vertex] (00) at  (8,0) {$0$};
	\node[vertex] (01) at  (11,0) {$1$};	
	
	\node[vertex] (10) at  (8,-3) {$0$};
	\node[vertex] (11) at  (11,-3) {$1$};	
	
	\node[vertex] (20) at  (8,-6) {$0$};
	\node[vertex] (21) at  (11,-6) {$1$};	
	
	\node[root vertex] (r) at  (9.5,-8) {};
	\node (rootLabel) at (9.5, -9) {$root$};
	\draw[edge, cross] (r) edge node[left]{$r'_0$} (20);
	\draw[edge, cross] (r) edge node[right]{$r'_1$} (21);

	\draw[edge] (10) edge node[left]{$p_{00}$}  (00);
	\draw[edge,cross] (11) edge node[left]{$p_{01}$}  (00);
	
	\draw[edge, cross] (10) edge node[right]{$p_{10}$} (01);
	\draw[edge] (11) edge node[right]{$p_{11}$} (01);
	
	\draw[edge] (20) edge node[left]{$c_{00}$} (10);
	\draw[edge, cross] (21) edge node[left]{$c_{01}$} (10);
	
	\draw[edge, cross] (20) edge node[right]{$c_{10}$} (11);
	\draw[edge] (21) edge node[right]{$c_{11}$} (11);

\end{tikzpicture}
\caption{shows QuDot Net before and after a SWAP operation}
\end{figure}

The last quantum gate needed to implement the QFT with QuDot Nets is the SWAP gate. QuDot Nets can efficiently implement the $n$-qubit SWAP gate where $n$ is the number of qubits in the quantum state. The $n$-qubit SWAP gate is not a collection of two qubit SWAP gates. The $n$-qubit SWAP gate operates on the entire state. QuDot Nets efficiently implement the SWAP gate by reversing the direction of all the edges in the QuDot Net and pointing the root node edges to the $n^{th}$ qubit layer. This is best illustrated in a diagram, see Figure 14

\newpage
\subsection{Quantum Fourier Transform}

\subsubsection{Scaling}

The QFT (and its inverse) is one of the most well studied algorithms in quantum computation. If the QFT is the last element of a quantum circuit then it is known as the \textit{terminal} QFT or \textit{semi-quantum} QFT and can be classically implemented [4]. QuDot Nets can also efficiently implement the semi-quantum QFT. We first re-write the QFT circuit in its semi-quantum form. Figure 15 shows the three qubit QFT. We use three qubits for the illustration

\begin{figure}[!ht]
\centering
\mbox{
\Qcircuit @C=1em @R=.7em {
	\lstick{\ket{x_1}} &  \gate{H} & \ctrl{1} & \ctrl{2} & \qw & \qw & \qw  & \qswap & \qw \\
	\lstick{\ket{x_2}} &  \qw & \gate{R2} & \qw &  \gate{H} & \ctrl{1} & \qw  & \qw \qwx  & \qw \\
	\lstick{\ket{x_3}} & \qw & \qw  & \gate{R3} & \qw  & \gate{R2} & \gate{H}  & \qswap \qwx & \qw
}
}
\caption{shows a 3 qubit Quantum Fourier Transform}
\end{figure}
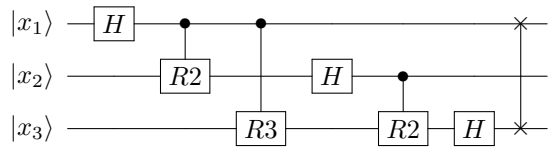

\vspace{5ex}
We now re-write the three qubit QFT in its semi-quantum form depicted in figure 16:

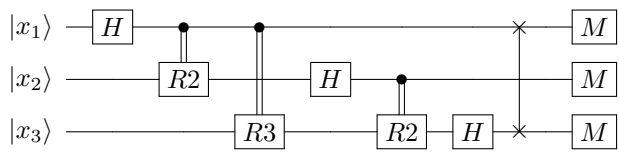
\begin{figure}[!ht]
\centering
\mbox{
\Qcircuit @C=1em @R=.7em {
	\lstick{\ket{x_1}} &  \gate{H} & \control \qw & \control \qw & \qw & \qw & \qw  & \qswap & \qw & \gate{M} \\
	\lstick{\ket{x_2}} &  \qw & \gate{R2} \cwx & \qw \cwx &  \gate{H} &\control \qw & \qw  & \qw \qwx  & \qw & \gate{M} \\
	\lstick{\ket{x_3}} & \qw & \qw  & \gate{R3} \cwx & \qw  & \gate{R2} \cwx & \gate{H}  & \qswap \qwx & \qw & \gate{M}
}
}
\caption{shows a 3 qubit Semi-Quantum Fourier Transform}
\end{figure}

\vspace{5ex}
Similarly, we rewrite the IQFT in its semi-quantum form:

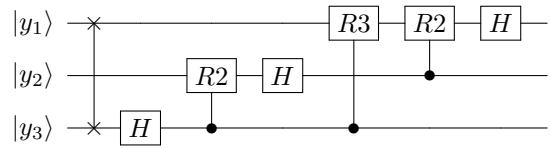
\begin{figure}[!ht]
\centering
\mbox{
\Qcircuit @C=1em @R=.7em {
	\lstick{\ket{y_1}} &  \qswap & \qw & \qw & \qw & \gate{R3} & \gate{R2} & \gate{H} & \qw \\
	\lstick{\ket{y_2}} &  \qw \qwx & \qw & \gate{R2} & \gate{H} & \qwx \qw & \control \qwx \qw & \qw & \qw  \\
	\lstick{\ket{y_3}} & \qswap \qwx & \gate{H} & \control \qwx \qw & \qw & \control \qwx \qw & \qw & \qw & \qw
}
}
\caption{shows a 3 qubit inverse Quantum Fourier Transform}
\end{figure}

\vspace{5ex}

\begin{figure}[!ht]
\centering
\mbox{
\Qcircuit @C=1em @R=.7em {
	\lstick{\ket{y_1}} &  \qswap & \qw & \qw & \qw & \gate{R3} & \gate{R2} & \gate{H} & \qw & \gate{M} \\
	\lstick{\ket{y_2}} &  \qw \qwx & \qw & \gate{R2} & \gate{H} & \cwx \qw & \control \cwx \qw & \qw & \qw & \gate{M} \\
	\lstick{\ket{y_3}} & \qswap \qwx & \gate{H} & \control \cwx \qw & \qw & \control \cwx \qw & \qw & \qw & \qw & \gate{M}
}
}
\caption{shows a 3 qubit Inverse Semi-Quantum Fourier Transform}
\end{figure}
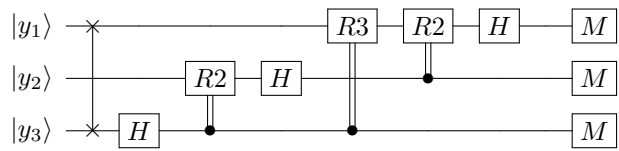

\vspace{5ex}
The QFT and IQFT require $n^2$ number of gates where $n$ is the number of qubits. An efficient implementation of the QFT and IQFT would require to scale as $O(n^2)$. We show quadratic scaling for the QFT and its inverse in our QuDot Net implementation for 6,000 qubits.

\begin{figure}[!h]
\includegraphics[scale=0.5]{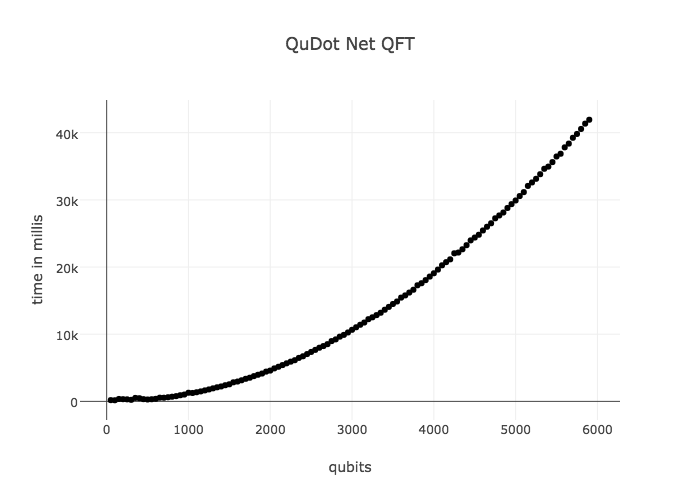}
\centering
\caption{shows quadratic scaling in QuDot Net execution of Quantum Fourier Transform}
\end{figure}

\begin{figure}[!h]
\includegraphics[scale=0.5]{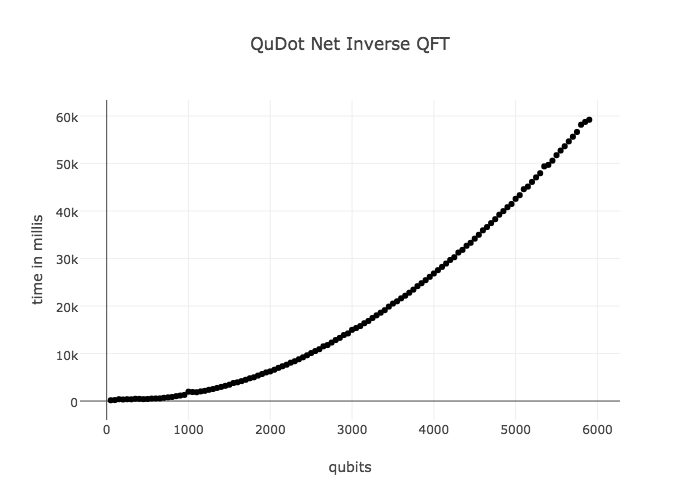}
\centering
\caption{shows quadratic scaling in QuDot Net execution of Inverse Quantum Fourier Transform}
\end{figure}

\subsubsection{Accuracy}

We test the accuracy of the QuDot Net implementation of the QFT by comparing to exact results against small four qubit test cases. We define six quantum states and show their expected probability distribution after the QFT is performed. We then run the QuDot Net implementation of the QFT on an ensemble of each state and perform a measurement. As the ensemble size increases for each state we observe that the expected probability distribution matches the observed probability distribution thus confirming the accuracy of the QuDot Net QFT implementation shown if figure 21.

$\ket{qft12}$ is obtained by applying $H$ to qubits 1 and 2. $\ket{qft24}$ is obtained by applying $H$ to qubits 2 and 4. The same mechanism is used to generate the other states. Expected probability distributions are shown in the Appendix. Figure 21 shows percent error of expected probability distribution vs measured probability distribution as ensemble size increases to 500,000 states.

\begin{enumerate}
	\item $\ket{qft12} = 0.5\ket{0000} + 0.5\ket{0100} + 0.5\ket{1000} + 0.5\ket{1100}$
	\item $\ket{qft13} = 0.5\ket{0000} + 0.5\ket{0010} + 0.5\ket{1000} + 0.5\ket{1010}$
	\item $\ket{qft14} = 0.5\ket{0000} + 0.5\ket{0001} + 0.5\ket{1000} + 0.5\ket{1001}$
	\item $\ket{qft23} = 0.5\ket{0000} + 0.5\ket{0010} + 0.5\ket{0100} + 0.5\ket{0110}$
	\item $\ket{qft24} = 0.5\ket{0000} + 0.5\ket{0001} + 0.5\ket{0100} + 0.5\ket{0101}$
	\item $\ket{qft34} = 0.5\ket{0000} + 0.5\ket{0001} + 0.5\ket{0010} + 0.5\ket{0011}$
\end{enumerate}

\begin{figure}[!h]
\includegraphics[scale=0.6]{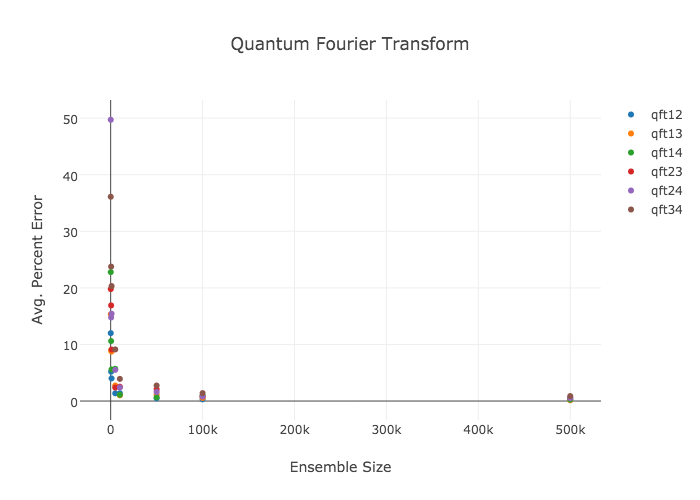}
\caption{shows exponential decrease in percent error as ensemble size grows}
\end{figure}

\section{Coherent Control Gates}

So far we have described semi-quantum computation which can be efficiently implemented using QuDot Nets. Implementing a Universal Quantum Computer (UQC) requires coherent control gates. Coherent control gates are control-$U$ gates when the control qubit is not measured. QuDot Nets can implement coherent control gates by using labeled, directed multi-digraphs. Once multiple labels are established single qubit gates operate on edge labels separately. Also, traversal must stay with the same edge labels. For example, suppose qubit 3 has edges with labels $a$ and $b$. If we apply an $H$ gate to qubit 3 then we apply the $H$ gate to all edges with label $a$ and then to all edges with label $b$ separately.

The efficiency of our implementation of coherent control-$U$ gates requires further study. Our results so far show that in the worst case the number of edges grows as $2^n$ where $n$ is the number of control qubits. This may explain why quantum computation cannot always provide an asymptotic increase for any algorithm. However, it remains to be seen if our multi-digraph implementation can be enhanced to remove duplicate edges and provide a speedup for the BQP class of algorithms such as Shor's Algorithm. If this is the case then QuDot Nets can show that BPP = BQP thus confirming the strong Church-Turing Thesis. It is worth noting that even certain specialized quantum hardware scales as $2^n$ for coherent control gates as explained by Martinis [5] and it is our position that coherent control gates will be the most difficult gates to scale with any custom hardware solution.

To apply a coherent control-$U$ gate where the control qubit is $c$ and the target qubit is $t$:
\begin{enumerate}
	\item go to qubit $c$
	\item for every incoming edge with label $l_i$ re-label as $l_i'$
	\item recursively add an edge with label $l_i'$ to every ancestor node with an edge label $l_i$
	\item for every outgoing edge with label $l_i$ re-label as $l_i'$
	\item recursively add an edge with label $l_i'$ to every descendant node with an edge label $l_i$
	\item go to qubit $t$
	\item apply the $U$ gate to the edges of qubit $t$ that have label $l_i'$
\end{enumerate}

For illustration purposes we will use colors as our labels but any unique identifier will do. Colors as labels should not be confused with edge coloring of a graph which will be used later in a different context. QuDot Nets alway start with a default label $l_1 = black$. We show an example of a coherent control gate by following the circuit in Figure 22.

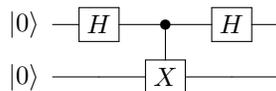
\begin{figure}[!h]
\centering
\mbox{
\Qcircuit @C=1em @R=.7em {
	\lstick{\ket{0}} &  \gate{H} & \ctrl{1} & \gate{H} & \qw \\
	\lstick{\ket{0}} & \qw & \gate{X} & \qw & \qw  \\
}
}
\caption{shows a simple quantum circuit used to explain coherent control gates}
\end{figure}

\newpage

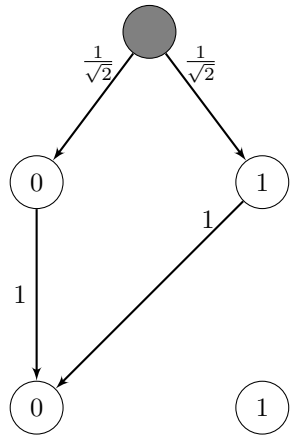
\begin{figure}[!ht]
\centering
\begin{tikzpicture}
	\node[root vertex] (r) at  (1.5,2) {};
	
	\node[vertex] (00) at  (0,0) {$0$};
	\node[vertex] (01) at  (3,0) {$1$};	
	
	\node[vertex] (10) at  (0,-3) {$0$};
	\node[vertex] (11) at  (3,-3) {$1$};	
	
	\draw[edge, cross] (r) edge node[left]{$\frac{1}{\sqrt{2}}$} (00);
	\draw[edge, cross] (r) edge node[right]{$\frac{1}{\sqrt{2}}$} (01);
	
	\draw[edge] (00) edge node[left]{$1$}  (10);
	\draw[edge, cross] (01) edge node[left]{$1$}  (10);

\end{tikzpicture}
\caption{After application of $H$ to first qubit}
\end{figure}

\begin{figure}[!ht]
\centering
\begin{tikzpicture}
	\node[root vertex] (r) at  (1.5,2) {};
	
	\node[vertex] (00) at  (0,0) {$0$};
	\node[vertex] (01) at  (3,0) {$1$};	
	
	\node[vertex] (10) at  (0,-3) {$0$};
	\node[vertex] (11) at  (3,-3) {$1$};	
	
	\draw[edge, cross] (r) edge node[left]{$\frac{1}{\sqrt{2}}$} (00);
	\draw[edge, cross, blue] (r) edge node[right]{$\frac{1}{\sqrt{2}}$} (01);
	
	\draw[edge] (00) edge node[left]{$1$}  (10);
	\draw[edge, cross, blue] (01) edge node[left]{$1$}  (10);

\end{tikzpicture}
\caption{After application of control to first qubit}
\end{figure}
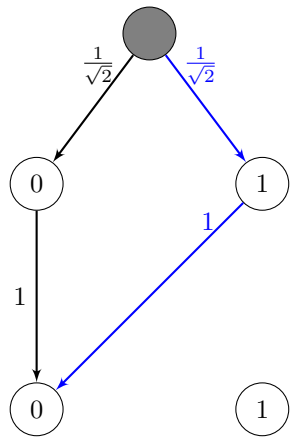

\begin{figure}[!ht]
\centering
\begin{tikzpicture}
	\node[root vertex] (r) at  (1.5,2) {};
	
	\node[vertex] (00) at  (0,0) {$0$};
	\node[vertex] (01) at  (3,0) {$1$};	
	
	\node[vertex] (10) at  (0,-3) {$0$};
	\node[vertex] (11) at  (3,-3) {$1$};	
	
	\draw[edge, cross] (r) edge node[left]{$\frac{1}{\sqrt{2}}$} (00);
	\draw[edge, cross, blue] (r) edge node[right]{$\frac{1}{\sqrt{2}}$} (01);
	
	\draw[edge] (00) edge node[left]{$1$}  (10);
	\draw[edge, blue] (01) edge node[right]{$1$}  (11);

\end{tikzpicture}
\caption{Now we apply the $X$ on the target only for the blue edges}
\end{figure}
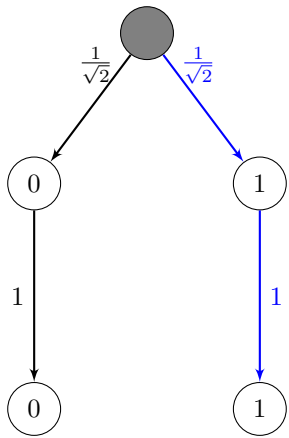

\newpage

\begin{figure}[!ht]
\centering
\begin{tikzpicture}
	\node[root vertex] (r) at  (1.5,2) {};
	
	\node[vertex] (00) at  (0,0) {$0$};
	\node[vertex] (01) at  (3,0) {$1$};	
	
	\node[vertex] (10) at  (0,-3) {$0$};
	\node[vertex] (11) at  (3,-3) {$1$};	
	
	\draw[edge, cross] (r) edge node[left]{$\frac{1}{2}$} (00);
	\draw[edge, blue, bend right] (r) edge node[left]{$\frac{1}{2}$} (00);
	
	\draw[edge,cross, blue] (r) edge node[right]{$\frac{-1}{2}$} (01);
	\draw[edge, bend left] (r) edge node[right]{$\frac{1}{2}$} (01);
	
	\draw[edge] (00) edge node[left]{$1$}  (10);
	\draw[edge, blue] (01) edge node[right]{$1$}  (11);
	
	\draw[edge, cross, blue] (00) edge node[right]{$1$} (11);
	\draw[edge, cross] (01) edge node[left]{$1$} (10);

\end{tikzpicture}
\caption{Finally we apply $H$ to the first qubit seperately for each edge label}
\end{figure}
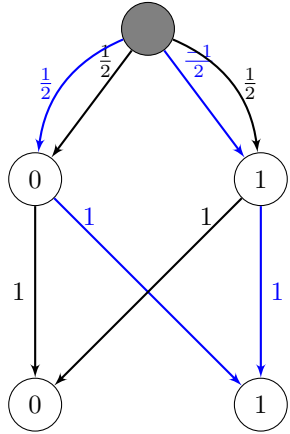

\vspace{4ex}

QuDot Net traversals must now stay on paths where edges have the same label. We have four possible traversals:

\begin{enumerate}
	\item $\ket{00}$
	\item $\ket{10}$
	\color{blue}
	\item $\ket{01}$
	\item $\ket{11}$
\end{enumerate}

QuDot Nets make it clear that control gates split the computational path. An interesting point of further study would be to see if quantum algorithms that achieve exponential speed up over their classical counterparts have a way to recombine certain computational paths which QuDot Nets represent as different labels on a multigraph.

\newpage
\section{Conclusive Remarks and Future Work}

We saw how the circuit model of quantum computation can be mapped to a special type of multi-digraph inspired by Bayesian Networks called a QuDot Net. With a QuDot Net, possible measurements are possible traversals where the probability amplitude is the product of each edge weight in the traversal. Single qubit gates can be implemented by providing rules for the manipulation of edge weights in a qubit layer. Two qubit control gates can be implemented by splitting the computational path making the QuDot Net a multi-digraph. We saw that semi-quantum computation can be efficiently implemented using QuDot Nets and we provided results for the efficient implementation of the QFT and Inverse QFT. 

What makes the quantum model of computation exponentially more powerful than the classical model of computation? This is still an open area of research. QuDot Nets show us that it is control gates that provide us with more power. A control gate splits the computational path and nature somehow keeps track of all the different edges. The most interesting question regarding QuDot Nets is ``Can we control the number of computational paths so they do not grow exponentially?". If we can then quantum computation can be efficiently implemented using classical hardware. If we cannot, can we at least approximate the answer using well know approximation techniques common in other NP-complete graph theoretic problems such as traveling salesmen and the knapsack problems? Even if there are errors in the approximations we still may be able to beat quantum hardware since quantum hardware also has errors but further work is needed to explore this possibility.

Einstein said many times ``God does not play dice with the world'' [8]. How would someone go about proving this? Einstein was making a precise computational statement $P=BQP$, however, computation complexity theory did not exist at the time. If we are able to show that $P=BQP$ then we confirm Einstein's statement. QuDot Nets may provide a way to show that $BPP=BQP$, that is, quantum computational randomness can be reduced to coin flipping. If we can control the exponential increase of the computational paths of QuDot Nets in bounded-error probabilistic polynomial time then we are a major step closer to proving Einstein's statement. Since there is much evidence suggesting that $P=BPP$ (which means randomized algorithms can be de-randomized) showing $BPP=BQP$ would take us  very close to $P=BQP$ which would have implications to quantum foundations.

General Relativity (GR) and Quantum Mechanics (QM) are the pillars of modern physics. However, we have not been able to unite them. Suppose we try to unite them, should we quantize GR or relativize QM? The physics community has been trying to do the later for the most part. Computationally, this takes the view that $P \ne BQP$. However, if we show that $P=BQP$ this implies that quantum mechanics can be de-randomized. Therefore, there is a deterministic theory that gives equivalent results as QM. Perhaps this theory is easier to unite with GR. Physics has historically progressed forward with unifications such as electricity and magnetism. Perhaps the next unification in physics is the unification of the computational complexity classes $BPP$ and $BQP$.

\newpage

\section{Appendix}

Table of expected probability distributions after applying QFT to test states used in Figure 21.

\begin{center}
	\begin{tabular}{| l | c | r |}
	\hline
	input state & possible measurement & measurement probability \\
	\hline
	$\ket{qft12}$ &  $\ket{0000}$ & 	0.24999999999999972 \\
	\hline
	& $\ket{0100}$ & 	0.24999999999999972 \\
	\hline
	& $\ket{1000}$ & 	0.24999999999999972 \\
	\hline
	&  $\ket{1100}$ & 	0.24999999999999972 \\
	\hline
         &  &  \\
	\hline
	
	$\ket{qft13}$ &  $\ket{0000}$ & 	0.24999999999999972  \\
	\hline
	& $\ket{0010}$ & 	0.12499999999999988 \\
	\hline
	& $\ket{0110}$ & 0.12499999999999985 \\
	\hline
	& $\ket{1000}$ & 	0.24999999999999972 \\
	\hline
	& $\ket{1010}$ & 	0.12499999999999988 \\
	\hline
	& $\ket{1110}$ & 0.12499999999999985 \\
	\hline
	&  &  \\
	\hline
	
	$\ket{qft14}$ &  $\ket{0000}$ & 	0.24999999999999972  \\
	\hline
	& $\ket{0010}$ & 0.21338834764831818 \\
	\hline
	& $\ket{0100}$ & 	0.12499999999999988 \\
	\hline
	& $\ket{0110}$ & 	0.03661165235168154 \\
	\hline
	& $\ket{1010}$ & 0.03661165235168150 \\
	\hline
	& $\ket{1100}$ & 0.12499999999999985 \\
	\hline
	& $\ket{1110}$ & 	0.21338834764831816 \\
	\hline
         &  &  \\
	\hline
	
	$\ket{qft23}$ &  $\ket{0000}$ & 	0.24999999999999972  \\
	\hline
	& $\ket{0001}$ & 0.10669417382415913 \\
	\hline
	& $\ket{0011}$ & 0.01830582617584076 \\
	\hline
	& $\ket{0101}$ & 0.01830582617584075 \\
	\hline
	& $\ket{0111}$ & 0.10669417382415908 \\
	\hline
	& $\ket{1000}$ & 0.24999999999999972 \\
	\hline
	& $\ket{1001}$ & 	0.10669417382415913 \\
	\hline
	& $\ket{1011}$ & 0.01830582617584076 \\
	\hline
	& $\ket{1101}$ & 0.01830582617584075 \\
	\hline
	& $\ket{1111}$ & 0.10669417382415908 \\
	\hline
         &  &  \\
	\hline
	$\ket{qft24}$ &  $\ket{0000}$ & 	0.24999999999999972  \\
	\hline
	& $\ket{0001}$ & 	0.12024247078195528 \\
	\hline
	& $\ket{0011}$ & 0.08641771452281801 \\
	\hline
	& $\ket{0100}$ & 	0.12499999999999988 \\
	\hline
	& $\ket{0101}$ & 0.03858228547718185 \\
	\hline
	& $\ket{0111}$ & 0.00475752921804457 \\
	\hline
	& $\ket{1001}$ & 	0.00475752921804457 \\
	\hline
	& $\ket{1011}$ & 0.03858228547718183 \\
	\hline
	& $\ket{1100}$ & 	0.12499999999999985 \\
	\hline
	& $\ket{1101}$ & 0.08641771452281801 \\
	\hline
	& $\ket{1111}$ & 0.12024247078195528 \\
	\hline
	\end{tabular}
\end{center}

\begin{center}
	\begin{tabular}{| l | c | r |}
	\hline
	input state & possible measurement & measurement probability \\
	\hline
	
	$\ket{qft34}$ &  $\ket{0000}$ & 	0.24999999999999972  \\
	\hline
	& $\ket{0001}$ & 0.20526673725850120 \\
	\hline
	& $\ket{0010}$ & 0.10669417382415913 \\
	\hline
	& $\ket{0011}$ & 0.02531116256909020 \\
	\hline
	& $\ket{0101}$ & 0.01130048978259131 \\
	\hline
	& $\ket{0110}$ & 	0.01830582617584077 \\
	\hline
	& $\ket{0111}$ & 0.00812161038981702 \\
	\hline
	& $\ket{1001}$ & 0.00812161038981702 \\
	\hline
	& $\ket{1010}$ & 0.01830582617584075\\
	\hline
	& $\ket{1011}$ & 0.01130048978259130 \\
	\hline
	& $\ket{1101}$ & 	0.02531116256909020 \\
	\hline
	& $\ket{1110}$ & 0.10669417382415908 \\
	\hline
	& $\ket{1111}$ & 	0.20526673725850114 \\
	\hline
	\end{tabular}
\end{center}

\subsection*{References}

\begin{enumerate}
	\item PW. Shore, SIAM Rev., \textbf{41(2)}, 303-332 (1999);
	\item P. Rebentrost, M. Mohseni, and S. Lloyd, Phys. Rev. Lett., \textbf{113}, 130503 (2014)
	\item RP. Feynman, International Journal of Theoretical Physics, \textbf{21(6)}, 467-488 (1982)
	\item RB. Griffiths and C. Niu, Phys. Rev. Lett. \textbf{76}, 3228 (1996)
	\item JM. Martinis, http://web.physics.ucsb.edu/~martinisgroup/papers/Martinis2012.pdf (2012)
	\item R. Jozsa, N Linden, Proceedings of the Royal Society A,  \textbf{459(2036)} (2003)
	\item E. Rieffel and W. Polak, MIT Press, 226 (2011)
	\item W. Hermanns, Branden Press, 58 (1983)
\end{enumerate}

\end{document}